\documentclass[12pt]{iopart}
\usepackage[dvips]{graphics}
\usepackage{epsfig}

 \newcommand\la{\langle}
 \newcommand\ra{\rangle}
 \newcommand\beq{\begin{equation}}
 
 \newcommand\eeq{\end{equation}}                                               
 \newcommand\beqn{\begin{eqnarray}}
 \newcommand\eeqn{\end{eqnarray}}

\def\mb{\,\mbox{mb}}
\def\fm{\,\mbox{fm}}
\def\GeV{\,\mbox{GeV}}
\def\Pom{{I\!\!P}}

\def\lsim{\mathrel{\rlap{\lower4pt\hbox{\hskip1pt$\sim$}}
    \raise1pt\hbox{$<$}}}         
\def\gsim{\mathrel{\rlap{\lower4pt\hbox{\hskip1pt$\sim$}}
    \raise1pt\hbox{$>$}}}

\begin{document} 
 
\title[Small Gluonic Spots in the Nucleon]
{Small Gluonic Spots in the Nucleon:\\
Searching for Signatures in Data}

\author{B.Z.~Kopeliovich$^{1,2}$ and B.~Povh$^{1}$} 

\address{$^{1}$Max-Planck-Institut f\"ur Kernphysik, Postfach 
103980, 69029 Heidelberg, Germany}
 
\address{$^{2}$Joint Institute for Nuclear Research, Dubna, 141980 
Moscow Region, Russia}

 \begin{abstract}
 Nuclear shadowing and color glass condensate are possible only at
sufficiently small $x$ where parton clouds of different nucleons overlap
in the longitudinal direction. Another condition vital for these effect,
an overlap of partons in impact parameters, is not easy to fulfill for
gluons which are located within small spots, as follows from the observed
weakness of diffractive gluon radiation (smallness of the triple-Pomeron
coupling). The predicted weakness of the leading twist gluon shadowing
has been confirmed recently by data for $J/\Psi$ production and Cronin
effect in $d-Au$ collisions at RHIC. Smallness of gluonic spots also
leads to a rather low value of $\alpha_{\Pom}^\prime$, the slope of the
Pomeron trajectory, confirmed by ZEUS data on elastic photoproduction of
$J/\Psi$. At the same time, saturation of unitarity for central $pp$
collisions leads to a substantial increase of $\alpha_{\Pom}^\prime$ in
good agreement with elastic $pp$ data.
 \end{abstract}

%\submitto{\JPG - Proceedings of Quark Matter 2004}
%\pacs{2485.+p, 12.38.Bx, 25.40.-h, 25.75.-q}

One of the most intriguing observations in soft hadronic collisions is
the smallness of the cross section of diffractive gluon radiation usually
expressed in terms of so called triple-Pomeron coupling, or the
Pomeron-proton total cross section. The dependence of the cross section
of the process $pp\to pX$ on the effective mass $M_X$ unambiguously
identifies the gluon radiation. Only partons with spin one can provide
$1/M_X^{2}$ behavior, while quarks lead to $1/M_X^{3}$. To see that this
diffractive excitation channel is indeed weak, one can compare the
measured Pomeron-proton cross section, $\sigma^{\Pom p}_{tot}\approx
2\mb$, with the expected $\sim50\mb$ corresponding to a meson-proton
cross section enlarged by the Casimir factor $9/4$, since the Pomeron is
expected to be a gluonic system. This estimate assumes that the gluon
dipole has a mean size of a meson, and the smallness of the observed
cross section signals that the gluonic system is much smaller. Adjusting
the mean gluon separation to data one arrives at rather small transverse
distance $r_0=0.3\fm$ \cite{kst2}.  This goes well along with earlier
observations of a small gluon correlation radius on the lattice and the
small instanton radius. Small gluonic spots in the proton were also
considered in the recent analysis \cite{sz}.

Of course, the spatial distribution of gluons in the proton is a 
fundamental issue and should affect many observables. 
Some of manifestation of this phenomenon in recent data from RHIC and 
HERA are considered below.

{\bf Gluon shadowing (saturation, color glass condensate).}

Gluons at small $x$ are less contracted in the longitudinal direction
than their sources. Therefore, at $x\lsim 1/(m_\pi R_A)$ gluons
originated from different nucleons can overlap and fuse leading to a
phenomenon known as shadowing. However, this is not a sufficient
condition for shadowing, the gluons must overlap in impact parameters 
as well. Apparently, smallness of the gluonic spots in the proton should
substantially reduce their transverse overlap. The mean number of gluons
overlapping with a given gluon in a nucleus can be estimated as,
 \beq
\la n_G\ra \sim \pi r_0^2 N \rho_A R_A\ ,
\label{10}
 \eeq
 where $N_G$ is the mean number of gluons in a proton. According to
estimates in \cite{k3p} the mean number of gluons $N_G\approx 2$ at the
energy of RHIC, and we arrive at a rather small number of overlaps, $\la
n_G\ra \sim 0.5$, quite far from what is needed for saturation. Such a
weak overlap also results in a weak gluon shadowing.  Predictions for
$x-$ and $Q^2$-dependences of nuclear shadowing can be found in
\cite{kst2}.

Note that gluon shadowing at low $Q^2$ corresponds to the triple-Pomeron
part of the inelastic Gribov's shadowing, and can be calculated directly
via diffraction cross section without any knowledge of the spatial
distribution of gluons in the proton. This shadowing turns out to be
quite weak \cite{kst2} and according to the dipole approach, the effect
may only decrease with $Q^2$. On the other hand, similar calculations
\cite{fs}, but with the triple-Pomeron part extracted from low statistics
diffractive data at HERA may lead to a very strong gluon shadowing.

\begin{figure}[thb] \centering 
\includegraphics[width=8cm]{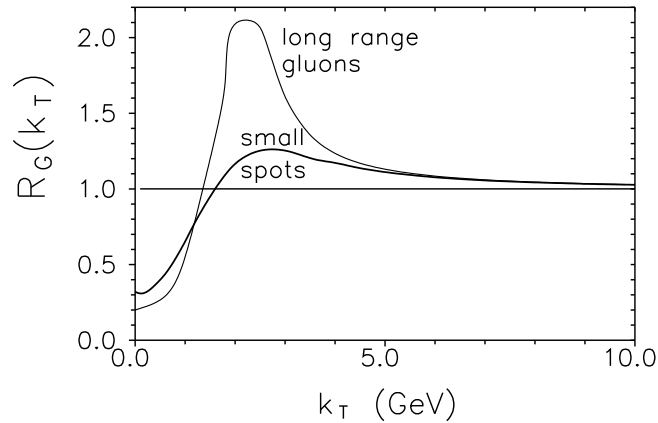} 
\caption{\label{ratio} 
 Ratio of $k_T$- distributions of gluons radiated in quark-nucleus
(A=200) and quark-nucleon collisions. The thick and thin curves
correspond to $r_0=0.3\fm$ and $1\fm$ respectively (see the text). }
 \end{figure}

So far we discussed the nuclear modification of gluon density, while the
transverse momentum distribution is also modified, the phenomenon which
nowadays is known as color class condensate \cite{mv}. Overlapping gluons
in nuclei are pushed to higher transverse momenta, and apparently this
effect should be also diminished if gluons are located within small
spots. In order to demonstrate this, we calculated the ratio of the $k_T$
distributions of gluons radiated by a quark colliding with a heavy
nucleus (A=200) and a proton. Both, the numerator and denominator are
calculated using the light-cone dipole formalism \cite{km,kst1} with the
light-cone quark-gluon wave function\cite{kst2},
 \beq
\Psi_{qG}(\vec r)=
\frac{2}{\pi}\,\sqrt{\frac{\alpha_s}{3}}\ 
\frac{\vec e\cdot\vec r}{r^2}\ 
\exp\left[-{r^2\over2r_0^2}\right]\ ,
\label{300}   
 \eeq
where $r_0$ is the parameter of the light-cone potential. We consider two 
possibilities, $r_0=0.3\fm$ corresponding to small gluonic spots, and 
$r_0=1\fm$ corresponding to a long-range propagation of gluons.
The results depicted in Fig.~\ref{ratio} 
 show how the magnitude of the Cronin effect is sensitive to the spatial
distribution of gluons in the proton. More realistic calculations
including initial state parton distributions and fragmentation functions
were done in \cite{knst}. Recent data for pion production in the PHENIX
experiment \cite{phenix} confirmed the predicted weak Cronin enhancement
(see also \cite{mine}) which is another confirmation for the spot
structure of the proton.

  {\bf Nuclear suppression of \boldmath$J/\Psi$ at RHIC.}

 The data for $J/\Psi$ production in $d-Au$ collisions released by the
PHENIX collaboration \cite{psi} provides probably the first direct access
to gluon shadowing. The whole nuclear suppression which also includes the
higher twist part related to the $\bar cc$ size \cite{kth}, turns out to
be quite weak in accordance with the prediction in \cite{kst2}, thus
confirming the idea of small gluonic spots. These data deserve few
comments:\\
 (i) The current nuclear-to-proton ratios must be diminished by about
$20\%$ due to the Gribov's and diffractive corrections missed in the
analysis (see details in \cite{mine}).  The central-to-peripheral ratio
is subject to even larger Gribov's corrections which are concentrated
mainly on the periphery of the collision \cite{mine}.\\
 (ii) Values of $x_2$ assigned to the data points in \cite{phenix} must
be increased. They should be related to the effective mass of the
produced $\bar cc$ pair which is heavier than $J/\Psi$. \\
 (iii) Comparison of the PHENIX and E866 data demonstrate a dramatic
violation of QCD factorization which is the basis for any current model
of $J/\Psi$ production. Although the above corrections bring the RHIC
data closer to E866, nevertheless a strong breakdown of the $x_2$ scaling
still remains. The reason lies in factorization breakdown towards the
kinematic bound of inclusive reactions \cite{bzk}. This is the case for
the minimal $x_2$ explored in the E866 experiment. In fact, there is no
room for gluon shadowing in the E866 data (on the contrary to the
conclusions in \cite{kth}), and the strong suppression observed at large
$x_F$ has a different origin \cite{bzk}.\\
 (iv) The prediction of a very weak leading twist gluon shadowing in
\cite{kst2} is rather reliable, since is dictated by diffractive data. At
the same time, a large higher twist correction to gluon shadowing was
predicted for $\chi_2$ production in \cite{kth}. This prediction is
rather model dependent, since is based on the potential model for gluon
interaction. Unfortunately, no data for $\chi_2$ production at RHIC are
available so far, while the available data \cite{psi} for $J/\Psi$ cannot
be used to test this expectation.

 {\bf Energy dependence of the elastic slope in \boldmath$J/\Psi$
photoproduction.}

Another manifestation of small gluonic spots in the proton is a rather
slow Gribov's diffusion which is responsible for the observed energy
dependence of the elastic slopes. Indeed, the mean square radius of a
spot is proportional to the product of $r_0^2$ and the mean number
of gluons in the spot. Correspondingly, it was predicted in \cite{k3p}
that the effective slope of the Pomeron trajectory reads,
 \beq
\alpha_{\Pom}^\prime = r_0^2\,\frac{\alpha_s}{3\pi}\,
\ln\left({s\over s_0}\right)\ .
\label{200}
 \eeq
 Different estimates of the running coupling $\alpha_s$ at the 
soft scale of the order of $1/r_0^2\sim 0.4\GeV^2$ converge at 
the value $\alpha_s\approx 0.4$ \cite{k3p}. With this value the Pomeron 
slope predicted in \cite{k3p} is rather small,
$\alpha_{\Pom}^\prime \approx 0.1\GeV^{-2}$.
 At first glance, this is in contradiction with the value
$\alpha_{\Pom}^\prime \approx 0.25\GeV^{-2}$ known from phenomenology of
soft hadronic collisions. However, a direct comparison with the energy
dependent slope $B_{pp}(s)$ of elastic $pp$ cross section demonstrates
excellent agreement \cite{k3p}. This is because the effective slope and
$\alpha_{\Pom}^\prime$ are substantially increased due to the effects of
unitarity saturation. Indeed, the $pp$ elastic amplitude Fourier
transformed to impact parameters nearly saturates unitarity for central
collisions where it is nearly independent of energy. All the observed
rise of the total cross section comes from peripheral collisions. This
leads to a fast growth of the interaction radius with energy.

At the same time, one expects almost no unitarity effects for hadronic
species having smaller cross sections, for instance $J/\Psi$-proton
elastic collisions. In this case the elastic amplitude is about order of
magnitude smaller than in $pp$, and the unitarity corrections are
negligible. Therefore, the observed value of $\alpha_{\Pom}^\prime$
should be close to the bare value Eq.~(\ref{200}). Indeed, the recent
measurements at HERA \cite{zeus}, have revealed a rather small value of
$\alpha_{\Pom}^\prime = 0.116 \pm 0.026\GeV^{-2}$ in a good accord with
the prediction Eq.~(\ref{200}) made in \cite{k3p}.  This is another
manifestation of small gluonic spots in the proton.

{\it Summarizing,} new data carrying the signatures of the small gluonic
spot structure of the proton have become available recently. This
includes the week gluon shadowing correction seeing in PHENIX data on
$J/\Psi$ production, a weak color glass condensate seeing in the Cronin
effect for pions, and the effective slope of the Pomeron trajectory in
elastic photoproduction of $J/\Psi$ which is twice as small as in elastic
$pp$ scattering.

\end{document}